\documentclass[a4paper,11pt]{article}
\pdfoutput=1 

\usepackage{jinstpub} 
\usepackage{pgf}                
\usepackage[normalem]{ulem}   
\usepackage{upgreek}
\usepackage[version=4]{mhchem}  
\usepackage{xspace}             
\newcommand{\rnzero}{$\ce{^{220}Rn}$\xspace}
\newcommand{\krm}{$\ce{^{83m}Kr}$\xspace}
\newcommand{\ambe}{$\ce{^{241}AmBe}$\xspace}

\newcommand{\roughly}{\sim\!\!}
\newcommand{\npmtamplification}{$(1.0 - 5.0)\times10^{6}~$}
\newcommand{\ncablelengthumbilical}{$12$ m~} 

\begin{document}

\title{The XENON1T Data Acquisition System}


\newcommand{\bern}{\affiliation{Albert Einstein Center for Fundamental Physics, University of Bern, 3012 Bern, Switzerland}}

\newcommand{\bologna}{\affiliation[4]{Department of Physics and Astronomy, University of Bologna and INFN-Bologna, 40126 Bologna, Italy}}

\newcommand{\chicago}{\affiliation[18]{Department of Physics \& Kavli Institute for Cosmological Physics, University of Chicago, Chicago, IL 60637, USA}}

\newcommand{\coimbra}{\affiliation[7]{LIBPhys, Department of Physics, University of Coimbra, 3004-516 Coimbra, Portugal}}

\newcommand{\columbia}{\affiliation[1]{Physics Department, Columbia University, New York, NY 10027, USA}}

\newcommand{\lngs}{\affiliation[16]{INFN-Laboratori Nazionali del Gran Sasso and Gran Sasso Science Institute, 67100 L'Aquila, Italy}}

\newcommand{\mainz}{\affiliation[5]{Institut f\"ur Physik \& Exzellenzcluster PRISMA, Johannes Gutenberg-Universit\"at Mainz, 55099 Mainz, Germany}}

\newcommand{\heidelberg}{\affiliation[11]{Max-Planck-Institut f\"ur Kernphysik, 69117 Heidelberg, Germany}}

\newcommand{\munster}{\affiliation[6]{Institut f\"ur Kernphysik, Westf\"alische Wilhelms-Universit\"at M\"unster, 48149 M\"unster, Germany}}

\newcommand{\nikhef}{\affiliation[3]{Nikhef and the University of Amsterdam, Science Park, 1098XG Amsterdam, Netherlands}}

\newcommand{\nyuad}{\affiliation[8]{New York University Abu Dhabi, Abu Dhabi, United Arab Emirates}}

\newcommand{\purdue}{\affiliation[21]{Department of Physics and Astronomy, Purdue University, West Lafayette, IN 47907, USA}}

\newcommand{\rpi}{\affiliation[10]{Department of Physics, Applied Physics and Astronomy, Rensselaer Polytechnic Institute, Troy, NY 12180, USA}}

\newcommand{\rice}{\affiliation[24]{Department of Physics and Astronomy, Rice University, Houston, TX 77005, USA}}

\newcommand{\stockholm}{\affiliation[2]{Oskar Klein Centre, Department of Physics, Stockholm University, AlbaNova, Stockholm SE-10691, Sweden}}

\newcommand{\subatech}{\affiliation[14]{SUBATECH, IMT Atlantique, CNRS/IN2P3, Universit\'e de Nantes, Nantes 44307, France}}

\newcommand{\torino}{\affiliation[17]{INFN-Torino and Osservatorio Astrofisico di Torino, 10125 Torino, Italy}}

\newcommand{\ucla}{\affiliation[25]{Physics \& Astronomy Department, University of California, Los Angeles, CA 90095, USA}}

\newcommand{\ucsd}{\affiliation[15]{Department of Physics, University of California, San Diego, CA 92093, USA}}

\newcommand{\wis}{\affiliation[12]{Department of Particle Physics and Astrophysics, Weizmann Institute of Science, Rehovot 7610001, Israel}}

\newcommand{\zurich}{\affiliation[9]{Physik-Institut, University of Zurich, 8057  Zurich, Switzerland}}

\newcommand{\paris}{\affiliation[22]{LPNHE, Universit\'{e} Pierre et Marie Curie, Universit\'{e} Paris Diderot, CNRS/IN2P3, Paris 75252, France}}

\newcommand{\freiburg}{\affiliation[13]{Physikalisches Institut, Universit\"at Freiburg, 79104 Freiburg, Germany}}

\newcommand{\lal}{\affiliation[23]{LAL, Universit\'e Paris-Sud, CNRS/IN2P3, Universit\'e Paris-Saclay, F-91405 Orsay, France}}

\newcommand{\naples}{\affiliation[19]{Department of Physics ``Ettore Pancini'', University of Napoli and INFN-Napoli, 80126 Napoli, Italy}}

\newcommand{\nagoya}{\affiliation[20]{Kobayashi-Maskawa Institute for the Origin of Particles and the Universe, Nagoya University, Furo-cho, Chikusa-ku, Nagoya, Aichi 464-8602, Japan}}

\author[1]{E.~Aprile,}\columbia
\author[2, 3]{J.~Aalbers,}\stockholm\nikhef
\author[4]{F.~Agostini,}\bologna
\author[5]{M.~Alfonsi,}\mainz
\author[6]{L.~Althueser,}\munster
\author[7]{F.~D.~Amaro,}\coimbra
\author[2]{V.~C.~Antochi,}
\author[8]{F.~Arneodo,}\nyuad
\author[2]{D.~Barge,}
\author[9]{L.~Baudis,}\zurich
\author[2]{B.~Bauermeister,}
\author[4]{L.~Bellagamba,} 
\author[8]{M.~L.~Benabderrahmane,}
\author[10]{T.~Berger,}\rpi
\author[3]{P.~A.~Breur,}
\author[9]{A.~Brown,}
\author[10]{E.~Brown,}
\author[11]{S.~Bruenner,}\heidelberg
\author[8]{G.~Bruno,}
\author[12]{R.~Budnik,}\wis
\author[13]{L.~B\"utikofer,}\freiburg
\author[9]{C.~Capelli,}
\author[7]{J.~M.~R.~Cardoso,}
\author[11]{D.~Cichon,}
\author[13]{D.~Coderre,}\emailAdd{daniel.coderre@physik.uni-freiburg.de}
\author[3, a]{A.~P.~Colijn,
\note[a]{Also at Institute for Subatomic Physics, Utrecht University, Utrecht, Netherlands}}
\author[2]{J.~Conrad,}
\author[14]{J.~P.~Cussonneau,}\subatech
\author[3]{M.~P.~Decowski,}
\author[1]{P.~de~Perio,}
\author[4]{P.~Di~Gangi,}
\author[8]{A.~Di~Giovanni,}
\author[14]{S.~Diglio,}
\author[13]{A.~Elykov,}
\author[11]{G.~Eurin,}
\author[15]{J.~Fei,}\ucsd 
% ...
\collaboration[1]{for the XENON Collaboration}
\collabconttrue
\collabtopfalse
\emailAdd{xenon@lngs.infn.it}

\newcommand{\caen}{\affiliation[26]{CAEN S.p.A. Viareggio, Lucca 55049, Italy }}

\exauthor{and }
\exauthor[26]{M.~Pieracci,}\caen
\exauthor[26]{C.~Tintori}

\abstract{The XENON1T liquid xenon time projection chamber is the most sensitive detector built to date for the measurement of direct interactions of weakly interacting massive particles with normal matter. The data acquisition system (DAQ) is constructed from commercial, open source, and custom components to digitize signals from the detector and store them for later analysis. The system achieves an extremely low signal threshold by triggering each channel independently, achieving a single photoelectron acceptance of $(93 \pm 3)\%$, and deferring the global trigger to a later, software stage. The event identification is based on MongoDB database queries and has over 98\% efficiency at recognizing interactions at the analysis threshold in the center of the target. A readout bandwidth over 300 MB/s is reached in calibration modes and is further expandable via parallelization. This DAQ system was successfully used during three years of operation of XENON1T.}


\arxivnumber{1906.00819} 

\maketitle
\flushbottom

\section{Overview}
\label{sec:intro}

The XENON1T experiment is the largest dual-phase liquid xenon time projection chamber (TPC) built to date. It was designed to measure nuclear recoils induced by dark matter in the form of weakly interacting massive particles (WIMPs). The detector has been in continuous operation at the INFN Laboratori Nazionali del Gran Sasso (LNGS) in Italy from its commissioning in 2016 until December 2018 and has produced the strongest exclusion limits to date on the coupling of WIMPs to matter \cite{xenon_1ty, xenon_sd, xenon_wp}. Particles incident on the 2.0 tonne liquid xenon target produce a prompt scintillation signal (S1) and delayed ionization signal, measured via proportional scintillation (S2), separated by a maximum time delay of 670~$\upmu$s. These signals are measured by two arrays of 127 and 121 Hamamatsu R11410-21 3" photomultiplier tubes (PMTs) \cite{xe_pmt} installed above and below the liquid xenon target. The energy region of interest for the standard spin-independent WIMP search is $[4.9-40.9]~\mathrm{keV}$ for nuclear recoils, but the bulk of background events, as well as regions of interest for other physics searches \cite{DEC}, are on scales up to a few MeV. The detector is installed in a large tank of deionized water instrumented as an active Cherenkov muon veto in order to reduce background and identify cosmic muons \cite{xe1t_muon_veto}. The XENON1T detector is described in detail in \cite{xe_instrumentation}.

\begin{figure}[htbp]
\centering 
\includegraphics[width=.98\textwidth]{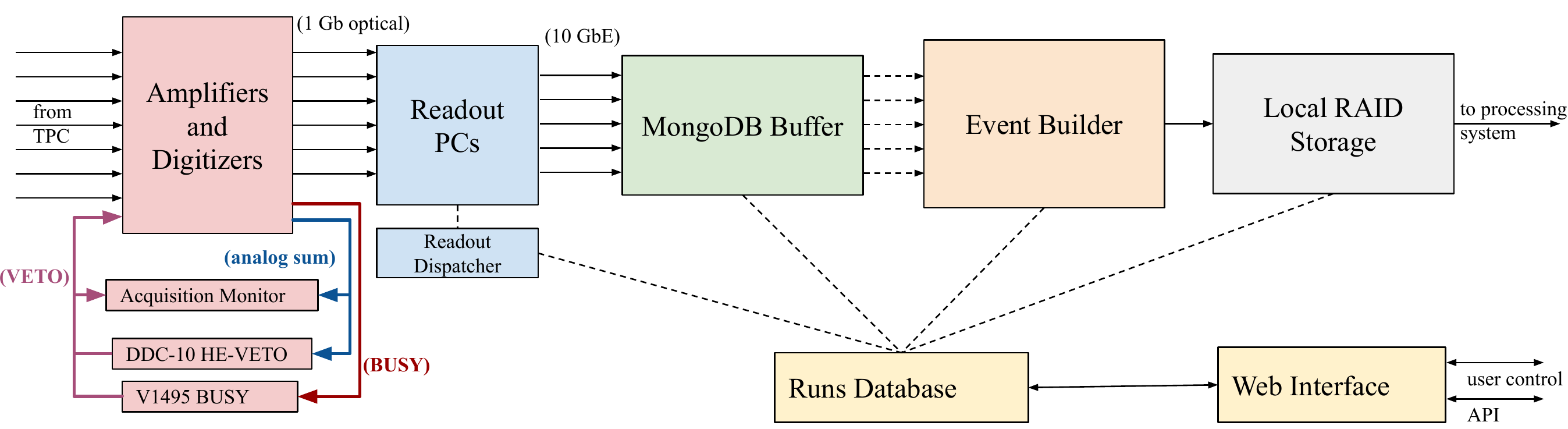}

\caption{\label{fig:diagram} A diagram of the XENON1T data acquisition system showing the main components. Signals from the TPC are digitized and reformatted for insertion into the MongoDB buffer.  
The event builder scans the buffer for events and writes them to storage.
The system is controlled via the run database backend, which is exposed to the user via a web interface. Details on each subsystem are explained in the text.}
\end{figure}

The data acquisition system (DAQ) is responsible for digitizing the signals from each PMT and identifying 
and recording every interaction with the target. The large variation in signal intensity  (keV up to MeV scale), as well as the two very different signal shapes of the S1 and S2 and the variable time delay between them, create a unique challenge for the DAQ. The DAQ is designed around a triggerless readout paradigm, where data is acquired constantly without a global trigger and physical interactions are identified later in software ("event building"). This allows for the lowest possible energy threshold, but necessitates custom readout logic to provide independent readout of each detector channel while maintaining full time synchronization. The system is centered around a NoSQL MongoDB database \cite{mongodb}, which provides fast storage, sorting, and retrieval of data. 
A software event builder queries this database asynchronously to identify interactions and write them to files. During normal operation the event builder runs within minutes of acquisition, but can be delayed for hours or days in case of disruptions further down the pipeline. The files are transferred to external sites and processed into a format useful for analysis by the XENON1T computing system \cite{pax_zenodo, benedikt}. A diagram of the DAQ is shown in Figure \ref{fig:diagram} and the subsystems are detailed in the following sections.

In dark matter search mode, the event rate identified by the event builder is about 5 Hz and the readout runs with <1\% deadtime, which is caused by long, periodic electrical noise emitted intermittently by individual channels. Nuclear recoil calibrations, which measure the detector response to a WIMP-like signal, are performed with an \ambe source or a deuterium-deuterium neutron generator \cite{xe1t_ng} placed in the water tank outside the cryostat and produce event rates of 10~Hz and 25~Hz, respectively. Other calibrations utilize radioactive sources mixed into the xenon target itself. \krm is injected once every few weeks for detector stability monitoring \cite{krm_cite}. \rnzero is used to measure the response to background-like low energy electronic recoil calibration events and is injected periodically throughout the science run \cite{xe100_rn220}. \krm and \rnzero calibrations have maximum event rates of 200~Hz and 100~Hz, respectively. Because XENON1T is a rare event search, the full digitized waveform of every event must be recorded for later scrutiny, necessitating a high throughput capability in calibration modes. In addition to these calibrations using radioactive sources, the PMT gains are calibrated with pulsed LEDs configured to provide signals at the single photoelectron level \cite{xe_pmt}. Table \ref{table1} shows an overview of the various operating modes as well as their maximum event rates and average event sizes.

\begin{table}[]
\centering
\caption{\label{table1} Event rates and sizes for various operating modes in XENON1T. For the internal sources \rnzero and \krm the event rate is highest right after the source is injected and decreases to the background rate over the course of several hours, at which point either another injection is performed or the calibration campaign is ended. The numbers with an asterisk indicate the event size after application of the hardware high energy veto, described in Section \ref{sec:electronics}.}

\begin{tabular}{|l|r|r|}
\hline
\textbf{Operation Mode}                & \textbf{Maximum Event Rate (Hz)} & \textbf{Average Event Size (MB)} \\ \hline
Dark matter search mode  & 6   & 1.68    \\ \hline
$^{220}\mathrm{Rn}$   & 100 & 1.66 (0.77*) \\ \hline
$^{83\mathrm{m}}\mathrm{Kr}$ & 200 & 0.33 \\ \hline
$^{241}\mathrm{AmBe}$ & 11 & 2.19 \\ \hline
Neutron Generator & 30 & 2.84 (1.02*) \\ \hline
LED & 450 & 0.30 \\ \hline
\end{tabular}

\end{table}


\section{Hardware and Readout}

The readout portion of the DAQ encompasses all components up to the MongoDB buffer in Figure~\ref{fig:diagram}. This stage is responsible for detecting signals from the TPC and putting them into a digital format suitable for software triggering.

\subsection{Frontend Electronics}\label{sec:electronics}

The electronics of XENON1T are housed in a counting room within a service building adjacent to the water tank containing the detector \cite{xe_instrumentation}. PMT high voltage is controlled via the XENON1T slow control system \cite{slowcontrol} and supplied by CAEN A1535N modules connected via custom-made passive low-pass filter boxes with a 2.8~MHz cutoff frequency to reduce electronic noise. The filter boxes reduce the RMS of the baseline from a mean of $(0.91\pm0.19)$~mV to $(0.31\pm0.06)$~mV, with the remaining dominant 25~kHz noise frequency originating from the high voltage power supplies themselves. Photons incident on the PMT photocathode are detected with an average quantum efficiency of 34.5\% and amplified by a channel-dependent gain of \npmtamplification \cite{xe_pmt}. The signals are routed by \ncablelengthumbilical PTFE-coated RG-196 coaxial cables \cite{xe_pmt} to the counting room, passing through potted vacuum feedthroughs (RHSeals), and connected to Phillips model 776 amplifiers \cite{phillips}. The amplifiers provide a fixed DC-coupled, non-inverting voltage gain of 10. One of the two outputs is connected to one of eight input channels on a CAEN model V1724 digitizer module in order to record signals from each PMT \cite{CAEN_V1724}. 31 of these modules are operated synchronously to record data from the TPC and one additional module records diagnostic data from the DAQ itself. One V1724 provides a 50~MHz clock synchronization signal that is propagated to all of the others such that the clock drift is constant across all digitizers. Acquisition start is synchronized with a NIM logic signal sent from a CAEN V2718 crate controller in order to  start the clock counters on all digitizers at the same instant. The V1724s digitize the input signals at a rate of 100~MHz with 2.25~V input range, 40~MHz input bandwidth, and 14~bit resolution. Data is read out from the digitizers by server-grade PCs via 7 optical links connected to CAEN A3818 PCIe cards with either 4 or 8 daisy-chained digitizers per link. Each optical link supports a readout bandwidth of up to 90~MB/s. By parallelizing readout over multiple optical links and PCs, a maximum readout rate of 300~MB/s is achieved for the complete system, which is sufficient to record the 100~Hz target rate for the \rnzero calibration mode. The maximum readout bandwidth is defined by the saturation point of the first optical link to reach its peak bandwidth of 90 MB/s, since beyond that point the DAQ will be capable of recording only partial events. Above about 100~Hz in \rnzero data, events begin to overlap due to long periods of activity following large S2 signals, during which many single electrons are extracted into the gas region \cite{sorensen_single_e}.  

\begin{figure}[htbp]
\centering 
\includegraphics[width=.9\textwidth]{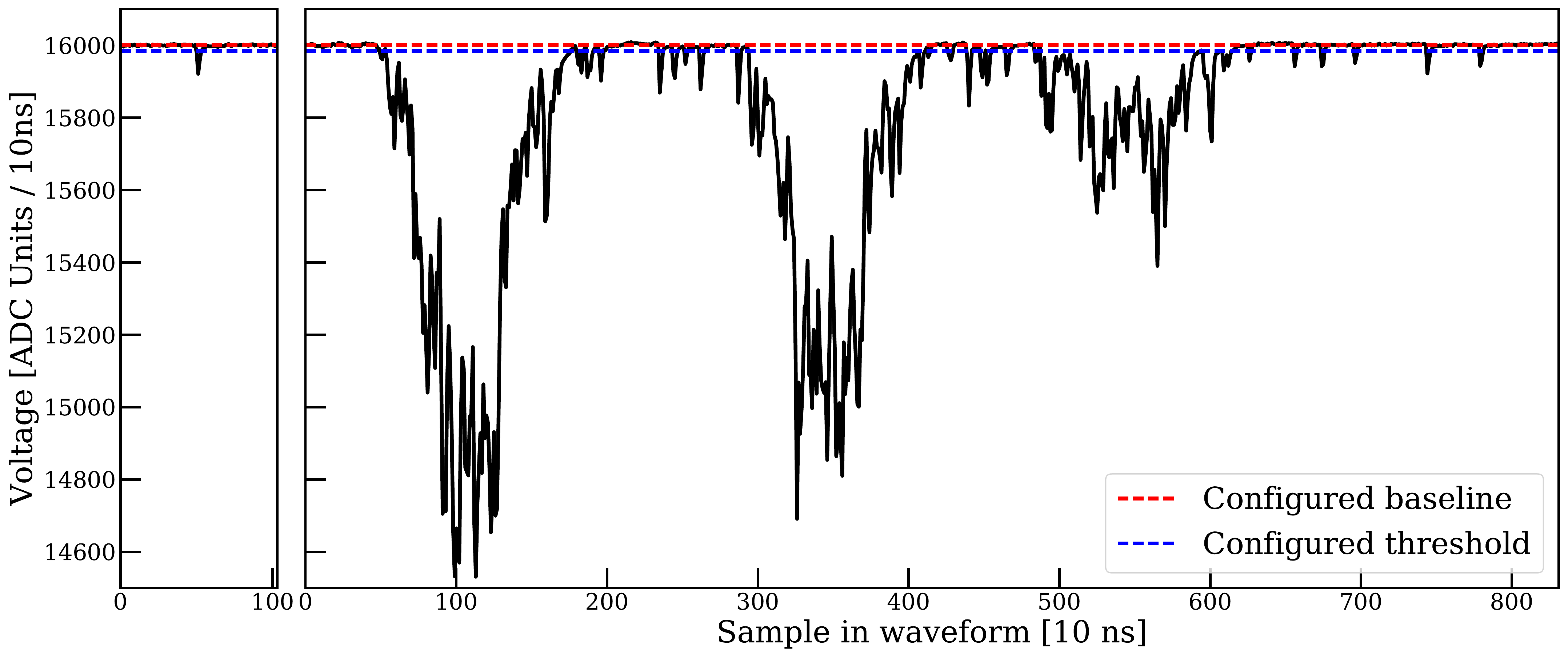}

\caption{\label{fig:pulses} Example self-trigger pulses from one PMT connected to a V1724 ADC. The left pulse is a single photoelectron (PE) and the right pulse is a long S2 candidate. The configured baseline (red) and self-trigger threshold (blue) are shown as dotted lines. The y-axis is zoomed to show detail, but the dynamic range of the digitizer extends to zero on this scale. The ADC scale is inverted, with the baseline at 16000 and the maximum (minimum) sample amplitude at zero (16383), where this range corresponds to the 2.25~V dynamic range of the digitizer.}
\end{figure}

Except during LED calibration measurements, which utilize an external periodic NIM trigger provided by a CAEN V2718 crate controller module,  digitizers are operated using a custom self-triggering firmware developed together with CAEN as a modification of their DPP-DAW firmware. It implements an independent threshold trigger on each channel ("self-trigger"), which stores and reads out the sampled PMT waveform of a single channel any time a signal on that channel exceeds a configurable threshold. The empty baseline data outside of the self-trigger window is not recorded. The firmware allows user configuration of the self-trigger threshold and the pre- and post-trigger acquisition windows. These windows are each set to 50 samples, corresponding to 500 ns, in order to provide enough empty baseline around each pulse for accurate fitting and subtraction. In case signals exceed the configured acquisition length, the window is automatically extended until the signal goes under threshold again for a user-configurable number of samples, allowing deadtime-less acquisition of both short, small S1 pulses and much larger and longer S2 pulses. This behavior is demonstrated in Figure \ref{fig:pulses}, which shows two self-triggered pulses. The pulse on the left axis is a single photoelectron (PE) and is the shortest length pulse possible (1~$\upmu$s), while the pulse on the right is a larger S2 that is recorded in its entirety by dynamically extending the acquisition window, in this case to more than 8~$\upmu$s. The maximum self-trigger length is defined by the size of the digitizers' internal memory and is over 3 ms. This is orders of magnitude longer than the longest pulses found in normal readout conditions during a XENON1T science run. The self-trigger threshold was set to 15 ADC counts, or 2.06 mV for the majority of channels, but is raised to a maximum of 30 ADC counts for 30 channels exhibiting abnormally high self-trigger rates due to periodic electronic noise from the high voltage power supplies or abnormal PMT behavior. The channel-dependent single PE acceptance averages to $(93 \pm 3)\%$ \cite{xenon_1ty}. 

\subsection{Readout Software}

Readout PCs run the kodiaq software package \cite{kodiaq_zenodo}, which reformats data from the binary CAEN format to MongoDB documents. Data is transferred from the digitizers to the readout PCs via block transfers mediated by the CAENVMElib C++ library. The readout software constantly polls the boards in round-robin fashion to read out any available new data. Upon completion of a block transfer, the transferred data is parsed into individual pulses. A \textit{pulse} refers to a single self-trigger from a single channel, which, together with its identifying meta-data, constitutes a single entry in the buffer database. During acquisition one software thread is dedicated to constantly polling the boards for new data in order to maximize readout bandwidth, while several other threads work to reformat the data as it arrives. MongoDB is a NoSQL database which organizes data into collections, analogous to tables in an SQL database. Single database entries in MongoDB are schema-less and are called documents (similar to SQL rows). Each document written by kodiaq contains the timestamp since run start in units of nanoseconds with 10~ns resolution, digitizer module number, channel number, compressed waveform samples as a binary field, and a MongoDB ObjectID field as a unique identifier. Kodiaq clients on different physical hosts (readout PCs) are controlled by a single \textit{dispatcher} software module, which interfaces with the user interface described later. The dispatcher monitors the state of the readout nodes, collects diagnostic information, sends commands, and creates and indexes collections in the buffer databases, which store data that is waiting to be triggered. In case one or more readout processes suffer a malfunction, the dispatcher stops the entire DAQ and propagates the error message to the user interface as well as to the XENON1T slow control system, where an alarm may be raised. 

The dispatcher is also responsible for interfacing with the runs database (see figure \ref{fig:diagram}) and creating metadata records for each run (see section \ref{sec:operation}). The dispatcher/slave architecture facilitates parallelization across several physical readout servers, enabling the system to reach a peak readout bandwidth of 300~MB/s. Communication between the dispatcher and slaves is performed via a custom protocol running over network sockets within the DAQ network. There is no hard limit on the level of parallelization possible and the system could in principle easily be extended by adding more hardware, with the eventual limit coming at the event building stage where all the independent readout pipelines must be combined by one process (see Section \ref{sec:trigger}). 

\subsection{Busy and High-Energy Veto Systems}

The CAEN V1724 digitizers have 8 MB onboard dual-ported memory per channel and will populate this buffer until the data are read by the readout software, at which point the memory is made available again. If the incoming data fills the buffer faster than it can be cleared, this memory can fill completely and the digitizer stops storing new events. In this case a \textit{busy} signal is raised by the digitizer on its front panel LVDS interface. The busy signals from all digitizers are propagated to a CAEN V1495 general purpose module where a custom FPGA firmware is configured to send a NIM logic signal in case any of the 31 V1724 data-transferring boards report a memory full condition. The NIM signal is fanned to an input on each board and the presence of this signal inhibits the board from recording new data. Therefore, if one board goes into the busy state, the data stream for all boards is effectively blocked until this state is lifted. This busy state is monitored via an additional digitizer, the \textit{acquisition monitor}, described in Section~\ref{sec:acqmon}. When the DAQ is operating using a hardware trigger, for example during LED calibrations and noise monitoring, the trigger NIM logic signal is passed through the V1495 module and inhibited in case any single board reports a busy state. This scheme ensures that for every recorded event, all data from all active channels is available. Partial events during which a busy condition was raised are removed from the later analysis using data from the acquisition monitor.

\begin{figure}[htbp]
\centering 
\includegraphics[width=.98\textwidth]{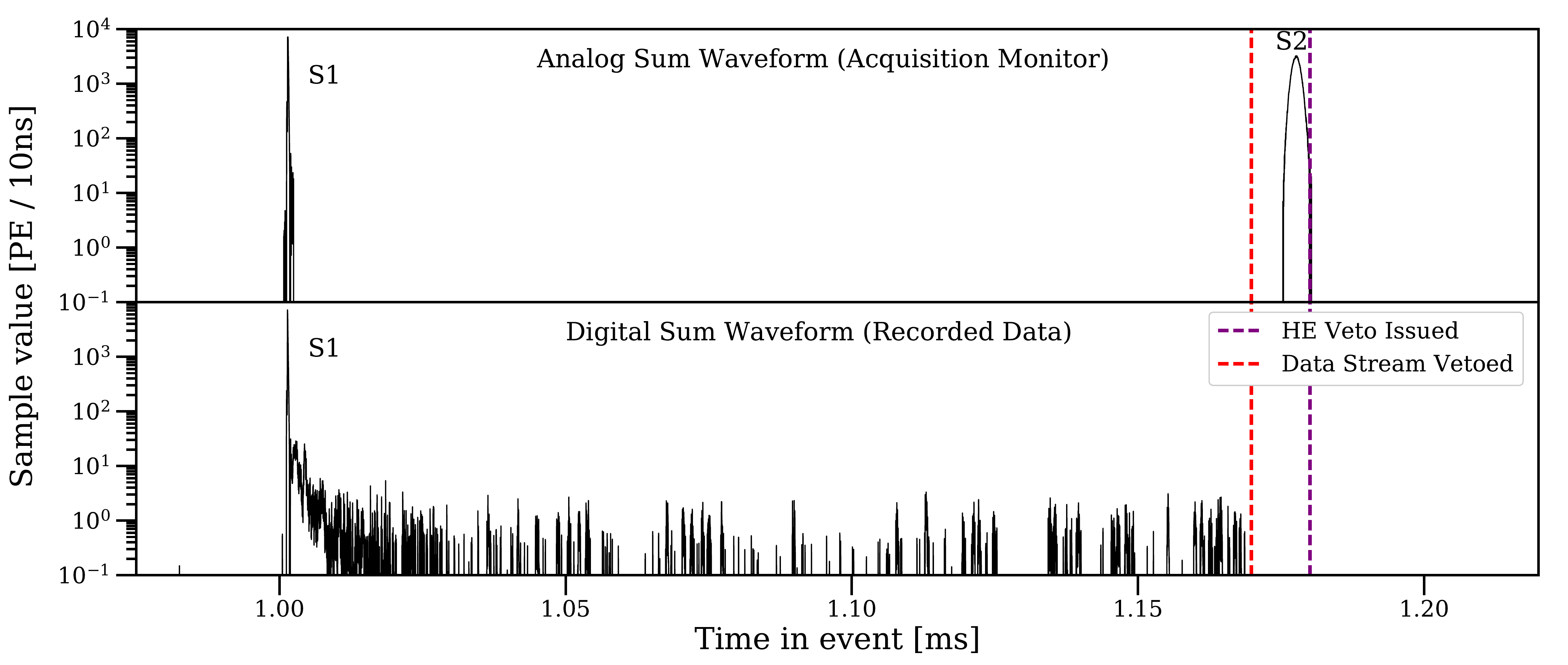}
\caption{\label{fig:hev_figure} An example event from $^{220}\mathrm{Rn}$ calibration removed by the high energy veto. The top panel shows the attenuated analog sum waveform from the bottom PMT array as recorded by the acquisition monitor and also used as the input to the DDC-10 high energy veto module. The large S2 fulfills the veto condition and a veto signal is issued after processing this peak (purple line). Due to the internal delay within the V1724, the data is vetoed already before the peak happens (red line), effectively removing it. The bottom panel is the digital sum waveform computed from the full-resolution recorded data, in which the S2 is no longer present. Note that the y-axis spans five orders of magnitude.}
\end{figure}

During neutron generator and  $^{220}\mathrm{Rn}$ calibration modes, the size of the data is greatly increased by high energy events outside the region of interest for low energy calibrations. In the case of $^{220}\mathrm{Rn}$ these are high energy $\alpha$-decay events and in the case of neutron generator data these are typically $\gamma$-decays following neutron capture processes. 
These high energy events are followed by long tails of delayed single electrons leading to a very high storage and bandwidth footprint and can cause unacceptably high dead-times, despite not being useful for most analyses. The (compressed) storage size in bytes of a XENON1T event is S2 dependent and is determined by data to be approximately
\begin{equation}\label{eq:datasize}
\mathrm{Event~Size} = 10~\mathrm{kB} + 4~\frac{\mathrm{B}}{\mathrm{PE}}\times\mathrm{S2}
\end{equation}
Here S2 represents the size of the S2 in PE. The constant term can be interpreted as the storage size required for digitizing an S1 and S2 plus the storage overhead, while the S2-dependent term represents the contribution of the post-S2 tail. There are multiple orders of magnitude in size difference between a typical low energy event ($\mathrm{S2}\sim10^2-10^3~\mathrm{PE}$) and a typical $\alpha$-decay event in \rnzero calibration data ($\mathrm{S2}~>~10^5~\mathrm{PE}$). In order to suppress these high energy events and preserve readout bandwidth for pertinent low-energy data, a veto system was developed based on a Skutek digital pulse processor DDC-10. It uses a BlackVME S6 motherboard hosting a Blackfin BF561 processor, programmable Spartan-6 LX150 FPGA, and a 10 channel 14-bit ADC card sampling at 100~MHz \cite{ddc10}. The input to this unit is the analog sum of the bottom array PMT channels, which is created using the second output of the Phillips 776 amplifiers and a series of Phillips 740 linear fan-in modules, the last two of which are modified with 30~dB attenuators to adjust the signal to be within the 2~V dynamic range of the DDC-10 digitizer input. A custom FPGA firmware on the high energy veto module constantly scans the input signal and searches for pulses with shapes compatible with S2s and integrated areas above a user-configurable threshold. This is set to approximately $1.5 \times 10^{5}$ PE, corresponding to several hundred keV. When a signal meeting these requirements is found, a NIM logic signal with a fixed length of 10~ms is generated and propagated to the digitizers using the same fan array used by the busy veto in order to inhibit acquisition of both the high-energy S2 and the most intense component of the subsequent single-electron tail. The storage size for one hour of $^{220}\mathrm{Rn}$ calibration data is about $150~\mathrm{GB}$ without or $35~\mathrm{GB}$ with the high energy veto active with no impact on the low energy calibration signals. Figure \ref{fig:hev_figure} illustrates the functionality of the high energy veto in an example event.

Both the high-energy and busy veto systems take advantage of another feature of the XENON custom CAEN V1724 digitizer firmware where data is not processed by the internal logic immediately but  delayed at the input stage in a FIFO by a user configurable time of up to 10.22~$\upmu$s (XENON1T uses this maximum value). Thus, a high energy S2 inducing a high-energy veto signal will be removed from the data stream by the veto signal that it prompted. This can be seen in Figure \ref{fig:hev_figure}, in which the veto signal is received in the data stream (red line) before it is actually issued in real time (purple line). 

\subsection{Acquisition Monitor}\label{sec:acqmon}

In addition to the 31 V1724 modules used to digitize the signals from the TPC PMTs, one additional V1724 module, the acquisition monitor, is used to digitize information about the DAQ system itself. This module is not connected to the busy module and is fed with a low enough rate of signals to ensure that it is never busy. In order to allow each channel to function independently, this digitizer is also equipped with the XENON custom firmware and operated in self-triggering mode. One channel records the attenuated analog sum of the bottom-array PMTs that is used for the high-energy veto input. If the DAQ does go busy due to high rate or a high-energy veto signal there is still a reduced level of information recorded from the detector during this time, as can be seen when comparing the top (sum signal recorded by the acquisition monitor) versus bottom (full resolution data) panels of Figure \ref{fig:hev_figure}. The other channels record the busy on/off, the high energy veto on/off, and the trigger from the muon veto system, which is used as an additional synchronization for the two systems (see section \ref{sec:mv}). The acquisition monitor data is stored in the main data stream and is available to analysts in high level data. The first level of selection conditions in the XENON1T analysis, which remove events where the busy, high-energy, or muon veto systems were active, relies on this data \cite{xe_analysis_paper_1}.

\subsection{Muon Veto Detector Readout}\label{sec:mv}

The water-Cherenkov muon veto detector surrounding the TPC consists of 84 Hamamatsu R5912-HQE PMTs, read out with a second DAQ system that uses a coincidence trigger module to induce acquisition in 11 CAEN V1724 digitizers via a hardware trigger \cite{xe1t_muon_veto}. Besides the different triggering paradigm, this system is nearly identical to the TPC system described in detail here and both systems share the same readout software, data flow framework, and analysis software. The data from the two detectors is combined at the analysis stage in order to remove events originating from a muon shower, especially muon-induced nuclear recoils.

The 50~MHz clock synchronization signal used by the TPC digitizers is also propagated to the muon veto modules to keep the clocks for both readout systems synchronized. In addition, a GPS-based absolute timing module is embedded into the muon veto system to provide time synchronization with potential external phenomena, such as galactic supernovae \cite{gps_timer}. A receiver module located on the laboratory surface computes the current time according to signals from GPS satellites and is connected by a fiber optic cable to a second, underground module located in the XENON counting room. Each time the muon veto is triggered, this module records and stores the current GPS time. In addition, a NIM logic signal is emitted every 15 seconds, which triggers the GPS module and is simultaneously recorded by the acquisition monitor. To compute the exact, global time for a given event, the DAQ retrieves the GPS time of the most recent muon veto or synchronization trigger and adds the time difference according to the digitizer clock. There is at most one bin (10~ns) of jitter over these few seconds, which is negligible for the purposes of this measurement. The DAQ for the upcoming XENONnT experiment will continue to use this method for time synchronization.

\section{Software Event Builder}\label{sec:trigger}

The continuous stream of data produced by the readout must be segmented into individual events for use in XENON1T's data processor PAX~\cite{pax_zenodo}. The data in between events, consisting primarily of PMT dark counts and  single electrons trailing high-energy events, is discarded to save storage space, except in special run modes for studying low-threshold, high-background processes.

The software trigger and event builder~\cite{jelle_thesis} perform this segmenting and pruning of the data stream. They read from the MongoDB buffer, search for events compatible with interactions in the detector, and produce archive files with events that are ready for further processing.

\subsection{MongoDB Buffer Databases}

The readout software, kodiaq, inserts data into three standalone MongoDB databases configured on three event builder computers. 
MongoDB's ``sharding" feature would have allowed operation of a single multi-host database, but it proved too slow (in various configurations) to maintain real-time insertion and query rates in calibration modes, likely due to overhead in maintenance of the combined index. Instead, the readout is configured such that a roughly equal amount of data is inserted into each of the three standalone databases by assigning one third of the digitizer modules to write to each database. In order to work around some aspects of the MongoDB internal memory management, most notably that the deletion of a document does not automatically free up the corresponding disk space, a new MongoDB collection is created every ~21 seconds on each machine and the data load is transfered to the newest collection. When the data in the collection is no longer needed, the entire collection is dropped, which deletes all data and indexes and frees up the corresponding storage space. The 21-second time window corresponds to the wraparound of the 31-bit, 100~MHz V1724 digitizer clock counter. All collections are indexed at initialization by timestamp, module identifier, and channel number, allowing for fast queries at the event builder stage. The trigger logic is able to detect pulses that span two collections by postponing processing of pulses appearing near the end of one collection until the next collection is read.

The database on each machine can use up to 128~GB RAM, 1~TB solid state storage, and several~TB hard disk space, allowing the DAQ to record data for days even if the connection to the file storage system is interrupted. During calibration data-taking, data is cleared by dropping each 21-second collection immediately after it is processed by the trigger, which typically happens less than two minutes after acquisition. During dark matter data-taking, pre-trigger data is stored in the buffer databases for six hours before being cleared. An external program monitors the Supernova Early Warning System (SNEWS) mail list and only flags data for purging if there is no supernova email warning \cite{snews}. This would enable the study of low-energy signals from supernova neutrinos, which may otherwise be partially below the trigger threshold \cite{lang_sn}.

\subsection{Trigger Logic and Performance}

The XENON1T software trigger logic is a single thread process included as part of the PAX data processor \cite{pax_zenodo}. It uses a rolling time window to query the MongoDB database and search for time-coincident pulses. Unpacking the full data payload of each pulse and combining this information would require a system with a prohibitively large amount of computing power, especially during high rate calibrations. Therefore, the payload (digitized waveform) of each self-trigger document is never examined, only the timestamp and channel number. These fields are indexed by the database, allowing the trigger to function in real time. The trigger logic is able to effectively identify S1 and S2 signals by recognizing the tight time coincidence (within 100~ns) of S1 signals compared to the broader distribution of S2 signals, which can be several $\upmu\mathrm{s}$ long. After classification as S1 or S2, a trigger decision is made based on the number of pulses that contribute to the signal, with the coincidence level set to 50 pulses for S1 and 60 pulses for S2.  

The low energy threshold of the XENON1T spin-(in)dependent WIMP search analyses \cite{xenon_1ty, xenon_sd, xenon_wp} is defined by a 3-fold coincidence requirement on the S1 signal. Due to the amplification of the S2 signal, interactions with S1s at this energy feature S2s that are orders of magnitude higher. Thus, it is the S2 trigger channel that is relevant for the low-energy performance of the DAQ. The threshold of 60 pulses eliminates triggers from one- or two-electron S2s, which are common following high-energy events, while achieving a high efficiency at the target analysis S2 threshold of 200 PE. Lower-energy signals can still be studied in special run modes where the trigger is disabled. Due to larger instrumental backgrounds at these low energies, only a small amount of such data is needed to get background-limited studies.

\begin{figure}[htbp]
    \centering 
    \includegraphics[width=\textwidth]{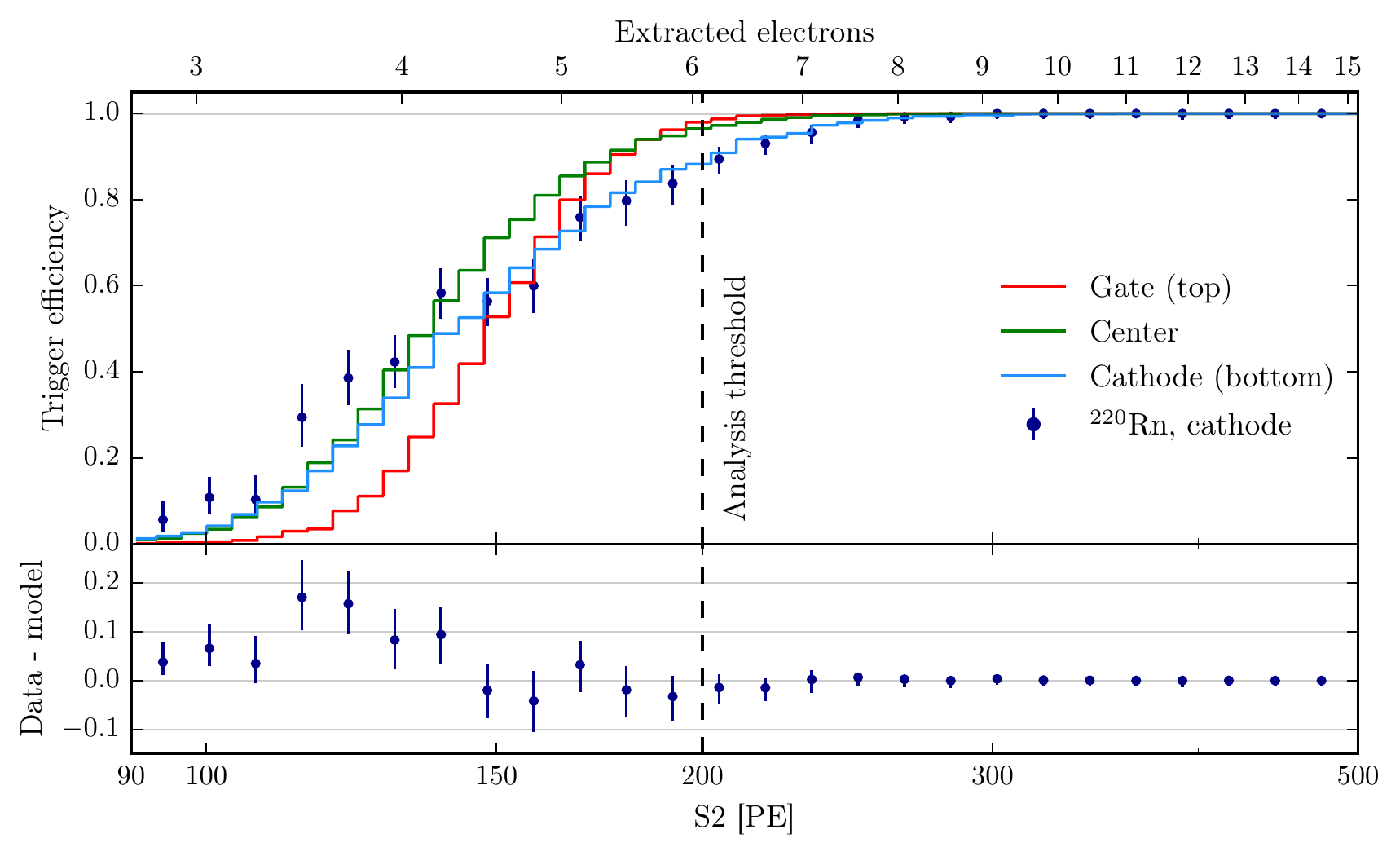}
    \caption{\label{fig:tr_eff} 
        Efficiency of the XENON1T S2 trigger versus observed S2 size in photoelectrons (PE, bottom x-axis) or equivalent number of extracted electrons (top x-axis). The colored lines show the efficiency determined from simulated waveforms, at different depths in the TPC. The dots show the S2 trigger efficiency, measured on S1-triggered events at the cathode in \rnzero calibration data, with statistical error bars. The lower panel shows this same efficiency measurement with the simulation-derived efficiency at the cathode subtracted.
        The trigger has a rapid activation between 100 and 150 PE and is >98\% efficient in the middle of the TPC above the analysis threshold of 200 PE (dashed line).
    }
\end{figure}

Figure \ref{fig:tr_eff} shows the trigger efficiency as a function of S2 size, as determined from simulated waveforms (curves) and measured using \rnzero calibration data events near the cathode (points). Due to large charge losses at the cathode (caused by the field inversion), events originating there can have S2s around the trigger threshold even if their S1s are large enough to induce a trigger. The S2 trigger efficiency is measured by the fraction of S1-triggered events for which the S2 also triggered. The dependency of the trigger efficiency on the depth of the interaction in the TPC is illustrated in Figure \ref{fig:tr_eff} for the top, middle, and bottom of the TPC. At the analysis threshold of 200 PE, the trigger accepts over 98\% of valid S2s in the center of the TPC.

\begin{figure}[htbp]
\centering 
\includegraphics[width=.98\textwidth]{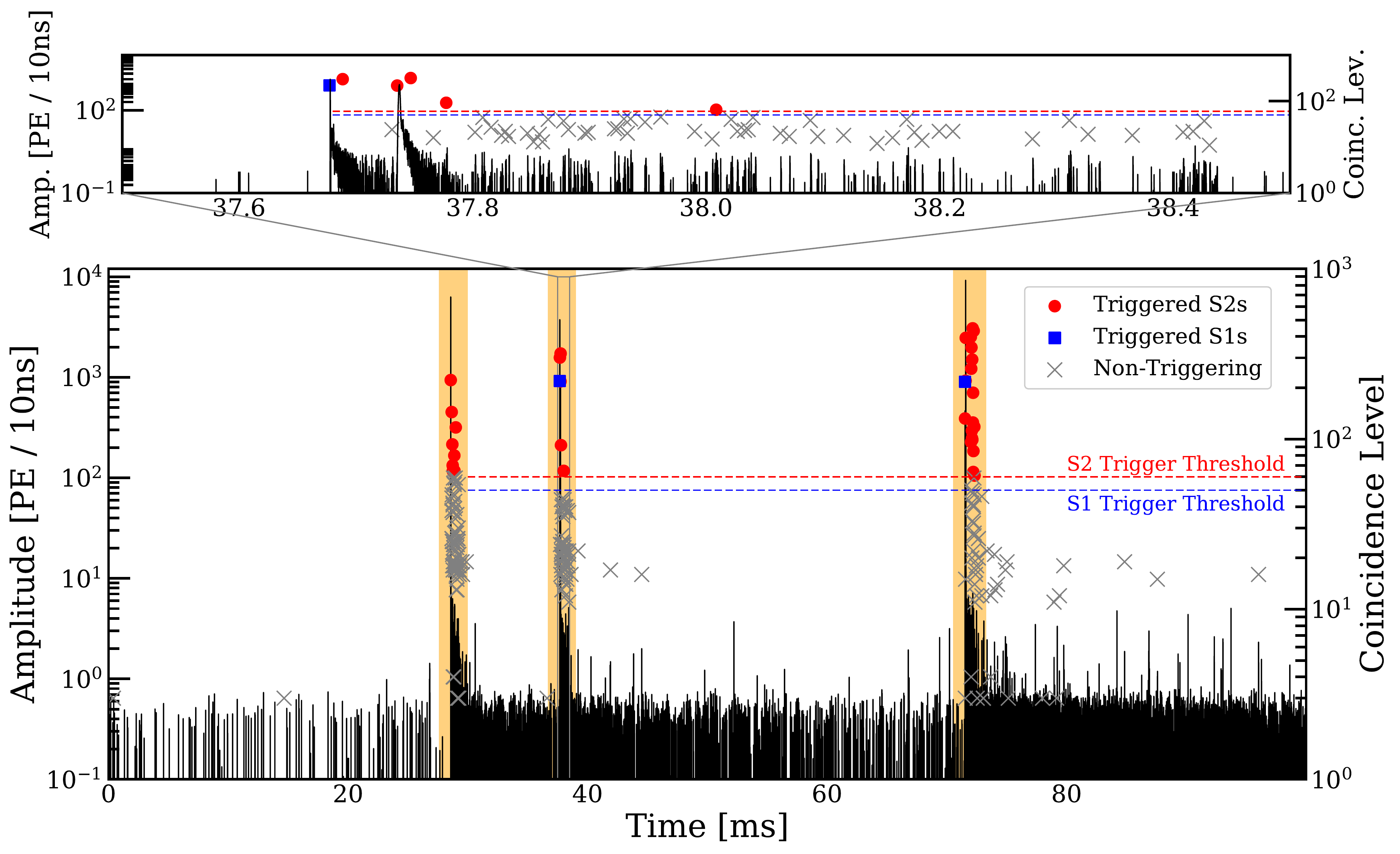}

\caption{\label{fig:tr_display} Sum waveform from 100~ms (about 14 full drift times) of XENON1T data at the input stage of the software trigger. S1 (blue) and S2 (red) peaks above the trigger threshold of 50 and 60 coincident signals, respectively, are marked and correspond to the right y-axis. Peaks detected below these coincidence thresholds (gray X markers) do not induce a trigger. The event windows are drawn as orange boxes and would be the only portion of this data written to disk during normal data-taking. The interesting portion of the middle event is shown in the upper panel.}
\end{figure}

A time window of $1~\mathrm{ms}$ is stored around each triggering S1 or S2 signal found above the trigger threshold, which was chosen to exceed the maximum drift time (670 $\upmu\mathrm{s}$) by several hundred $\upmu\mathrm{s}$ in order to always record some time before and after each event. Overlapping event windows are combined up to a maximum event size of 10 ms, at which point they are artificially truncated. This truncation causes a short amount of dead time, but since it only takes place for events that are over 15 drift lengths long, it is irrelevant for data acquired under normal conditions. The operating principle of the software trigger is presented in Figure \ref{fig:tr_display}.

The output of the event builder is a series of start and end times corresponding to event ranges. These are sent to a file-building system, which queries all pulses corresponding to the event time ranges, reformats them to the storage format, and writes out each event. This system consists of $\roughly20$ worker processes on each of the three PCs that also host the MongoDB buffer databases, controlled via two \emph{RabbitMQ} task queues~\cite{rabbitmq}. The event builder puts event time ranges on a first queue, from which the workers read. The workers put the finished event objects in a second queue. Finally, a single process reads the event objects, then streams them to file in the order in which the events were acquired. At the XENON1T background event rate of $\roughly5~\mathrm{Hz}$, the output data stream is approximately 30~GB/hour. During calibration, output rates up to a 250~GB/hour are possible for data collected without the high energy veto active. 

\section{Operation and Control}

Besides the main processes reading and triggering data, various support systems are used to combine these processes into a working, controllable system and to keep track of the data collected. This section describes the runs database and web interface, shown as yellow boxes in Figure \ref{fig:diagram}.

\subsection{Run Database}\label{sec:operation}

The communication backend of the DAQ, and also of the data processing system not described here (see \cite{benedikt}), is another MongoDB database called the runs database. This is a three-member replica set with the primary at LNGS and two secondaries located at the University of Chicago and the University of Stockholm, which are the main XENON1T computing sites. Locating the replicas at the remote computing sites allows for local read access from computer nodes and seamless operation of many components of the processing and analysis systems independent of the LNGS network. Data recorded by the DAQ is organized into data sets called \emph{runs}, where each run corresponds to data acquired for a given consecutive range of time (usually one hour) under identical settings. The runs database holds a record of all DAQ, trigger, and data processor settings, important operational conditions, operator comments, and an index of all locations where any type of data is stored for each run of the DAQ. Metadata for muon veto runs are also stored in this database in the same way. This index is used within the DAQ system for subsystem communication. The digitizer readout code, kodiaq, makes an entry at run start that it wrote untriggered data to the MongoDB buffer. The event builder registers this and begins triggering the run, while also making an entry that it began writing a data file to the local storage array at LNGS. The transfer system registers this and transfers the written raw data out to the storage and processing sites, and so on. The computing system uses the runs database as an index to track the location of all data associated with a particular run. For example, one run may have raw data written to tape archive, one to two additional copies of raw data in cloud storage, and one or more copies of processed data at analysis sites, all of which is tracked in the run document. Access to this database is also embedded into the XENON1T analysis software and analysts query the run database to select data with certain properties, for example data acquired with a specific calibration source or acquired with certain operating conditions. 

\subsection{User Interfaces}

The DAQ system is accessed by users via a web interface with two entry points: the graphical interface, which is written in HTML and JavaScript for human interaction, and the API, which is for programmatic access. The backend is powered by the Django framework \cite{django}. The human user interface allows configuration of the DAQ by experts into pre-defined readout modes that define the configuration for all parts of the system for calibration, dark matter search, and other data-taking. Both the TPC and muon veto subdetectors are controlled over this interface either independently or in synchronous operation. Operators can monitor the data rates, block transfer rates, and memory and CPU usage of each readout node. The software trigger also provides diagnostic information including its status, database buffer storage capacity, current run, and current event rate. A monitoring app allows inspection of the lone and coincident self-trigger rate per channel for any run and a live oscilloscope displays the most recent pulses recorded in any channel at 10~ns sampling rate. Both of these have proven instrumental for immediate remote diagnosis of detector anomalies. 

\begin{figure}[htbp]
\centering 
\includegraphics[width=.98\textwidth]{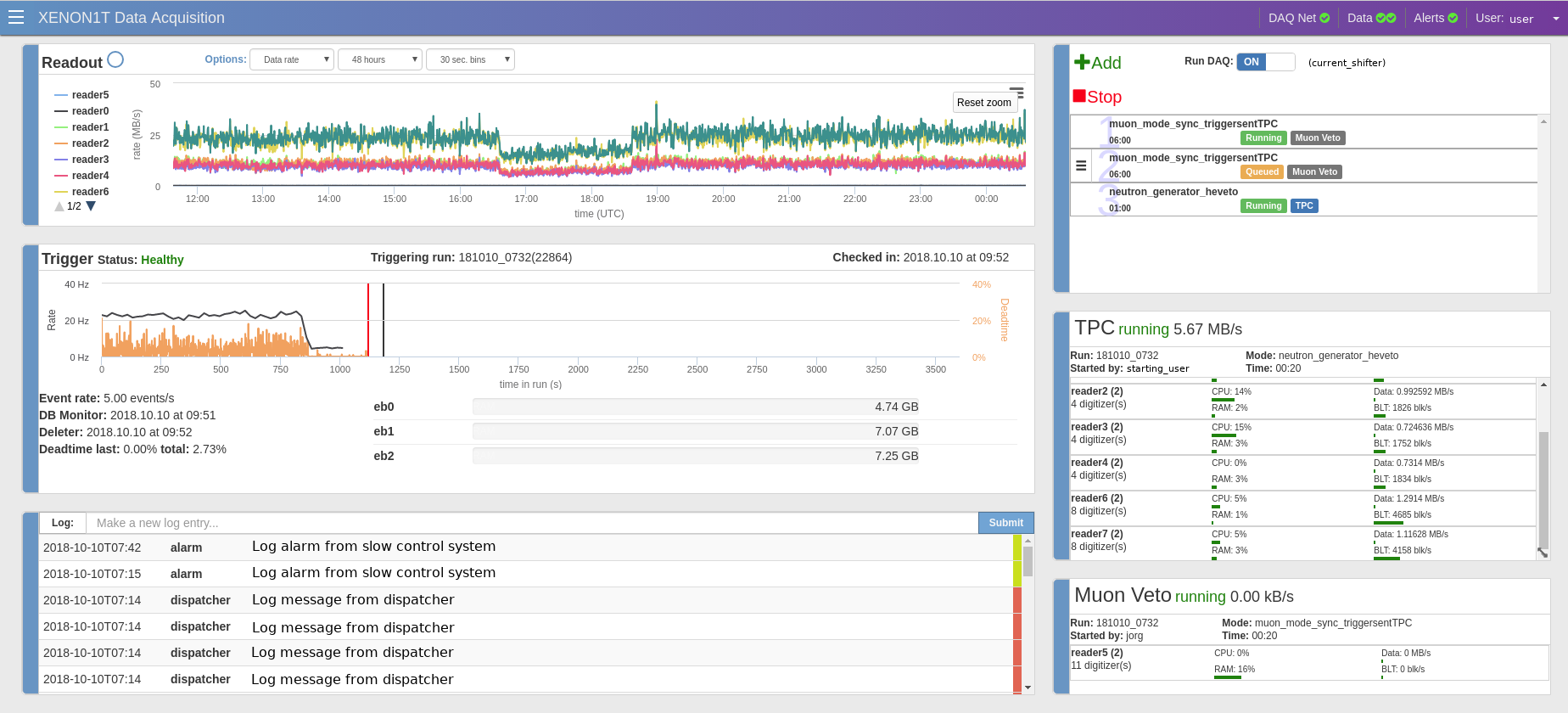}
\caption{\label{fig:screenshot} One of the views on the DAQ user interface allowing for control of the system. Clockwise from top left: (1) transfer rate history for readout nodes, (2) run control panel allowing data-taking to be started and stopped with various settings, current readout status for (3) TPC and (4) muon veto processes, (5) DAQ log messages, (6) software trigger status showing event rate, dead time, and pre-trigger buffer occupancy. }
\end{figure}

A graphical interface to the runs database allows operators to apply single-word tags to mark exceptional runs (e.g., "test" for test runs not intended to be used in the dark matter exposure), add extended explanatory comments associated with single runs, view the processing status and data locations of all data types associated with a run, and browse and search runs by their properties. The interface also provides general logging functionality, user control, and an organizational platform for operation shifts. Figure \ref{fig:screenshot} shows a screen shot from the control page of the web interface.

The API is implemented with the Django tastypie framework \cite{django-tastypie} and provides programmatic access to the runs database and DAQ backend. This API is world-accessible and is used by the computing and data transfer system to read and record processing statuses and storage locations for individual runs. The slow control system uses this API to query the DAQ status and monitor the system. Key parameters, including the data transfer rate, are also stored in the slow control databases. Disruptions within the DAQ are also propagated to the slow control system over the API, and alarms are raised in the case of DAQ errors, loss of connectivity to the API, or if the data rate has been zero for an extended time. The slow control alarm system then alerts operators and experts via email and telephone \cite{slowcontrol}.

\section{Conclusion}

The XENON1T DAQ system takes advantage of the speed and data-structure of NoSQL databases to implement a flexible software trigger. At low data rates such as in dark matter search mode, this allows the trigger decision to be deferred, possibly indefinitely, or performed remotely. The flexible software trigger is made possible by the use of a custom digitizer firmware to stream all data from the detector to software without any reduction. Every photon measured by the PMTs is inserted as a document in the MongoDB buffer database and used to identify events.

Additionally, the use of a modern web application to control the experiment allows a greater degree of operational flexibility and has allowed operators to forego a conventional control center. Touch-screen tablets allow full control of the system and inform operators of disruption, day or night. This system is the precursor to another iteration, currently being designed and commissioned at the XENONnT experiment, which will further expand upon the strengths of the XENON1T DAQ by forgoing the trigger stage completely in favor of fast, online data processing \cite{MC_paper}.

\end{document}